# Unidirectional Raman emissions of Stokes photons via chiral atom-photon coupling in a ring cavity

Haole Jiao[1,2], Minjie Wang[1,2], Jiajin Lu[1,2], Can Sun[1,2], Zhifang Yang[1,2], Mengqi Xi[1,2], Shujing Li,[1,2] and Hai Wang*,[1,2]

*1 The State Key Laboratory of Quantum Optics and Quantum Optics Devices, Institute of Opto-Electronics,*
*Shanxi University, Taiyuan 030006, China*
*2 Collaborative Innovation Center of Extreme Optics, Shanxi University, Taiyuan 030006, China*

The non-reciprocal (unidirectional) atom-photon couplings are crucial for modern photonics ranging from chiral quantum networks to cold-atom many-body physics. In the presented experiment, we demonstrated unidirectional Raman emission of Stokes photons from $^{87}$Rb atoms in a ring cavity. A bias magnetic field $B_0$ is applied along z-direction on the atoms to define the quantum axis, which breaks the time inverse symmetry. By transversely applying write laser pulses to drive a π-transition of the atoms, we generate spontaneous Raman emissions of Stokes photons from a chiral ($\sigma^+$) transition. The emissions are coupled into the clock-wise (z-direction) and counter-clock-wise (-z-direction) modes of a running-wave cavity, respectively. According to the mirror (parity) symmetry of the atom-light coupling, we demonstrated that spins (polarizations) of the Stokes fields are correlated with their propagation directions along z and –z -axis. The Stokes emissions constrained to the spin-momentum correlation are found to be violation of Kirchhoff's law of thermal radiation. Based on the correlation, we demonstrated that the Stokes emissions propagate along the clock-wise or counter-clock-wise via polarization dissipation. The directional factor is up to 1500:1.

# 1. Introduction

A long-term task in optical quantum information science is to develop techniques that can enable light-matter interaction in non-reciprocal ways.[1,2] Chiral quantum light-matter interfaces promise non-reciprocal coupling (NRC) of quantum emitters (atoms) with photons, where, the atoms emit (scattering, absorb) photons into forward and backward directions in an asymmetric way.[1] Based on the chiral quantum interfaces, the schemes of[2-4] chiral quantum networks or the simulations of many body physics[5] have been proposed. Recently, the unidirectional (forward- or backward- direction) light emissions from chiral (circularly-polarized) transitions of a quantum emitter, which is set closed to or embedded in nanophotonic waveguides, have been demonstrated.[6-10] The theoretical studies on the directionality of photon transport in waveguides have been reported.[11,12] Additionally, quantum optical circulator has been demonstrated with the system built by a single chiral atom coupled to whispering-gallery-mode (WGM) microresonators. These demonstrations are based on the locking (correlation) between spin and propagation direction of light fields that are transversely confined by nanophotonic devices,[1, 13] which has been pointed out to be a quantum spin Hall effect of light.[14] However, in these schemes, the quantum emitters are set close to dielectric surfaces, which perturb the strong atom-photon interacting.[4]

The couplings (interactions) of cold and ultra-cold atomic ensembles with light fields can be effectively enhanced by optical cavities and may reach strong-coupling level with high-finesse cavity.[15] In the past two decades, atom-cavity systems are used as important tools to investigate quantum information science,[16-22] superradiant dynamics of atoms,[22-25] and single-photon-level isolators.[26,27] The chiral interactions between matter and light relate with the nonreciprocal emission, absorption, and transmission of light.[1] Recently, nonreciprocal thermal photonics have been widely demonstrated with a magneto-optical material.[28-31] It has been pointed out that the chiral transitions of atoms are effectively coupled with circularly-polarized light.[32] However, the correlation between the spins (polarizations) and propagations directions of the light emission from a chiral transition of atoms in free space, remained elusive, so far.

In the presented experiment, we demonstrated unidirectional emission of Stokes photons from a chiral transition of Rb atoms coupled with a ring cavity. Such a demonstration is based on the spin-direction correlation, which refers to that the Stokes emissions are right-hand-circular polarization when they propagate along +z-direction and are left-hand-circular when –z direction, respectively. This spin-direction (spin-momentum) correlation can be known from the left-right (parity) symmetry.[33] Constrained to the spin-momentum correlation, the Stokes emissions from a chiral transition are found to be a violation of Kirchhoff's law of thermal radiation. The measured polarization purity of any one of the propagating direction along the clock-wise (CW) or counter-clock-wise (CCW) mode is up to ~1500:1. Based on the spin-momentum correlation, we demonstrated an asymmetrical Stokes emission along the CW and CCW modes of the ring cavity via polarization dissipating, with a relative directional factor $\beta$ of ~1500:1.

## 2. Polarization-direction correlation

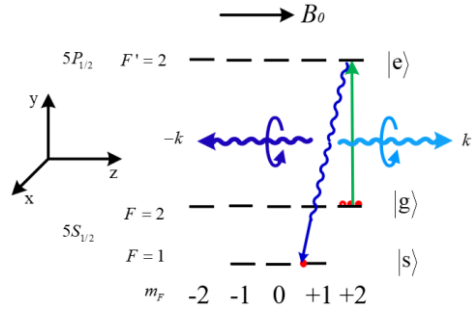

**Figure 1.** the relative atomic levels used for the Stokes emissions.

**Figure 1.** shows the relative atomic levels. A bias magnetic field $B_0 = 4G$ is applied on the atoms along z-axis to define quantum axis of the atoms. The atoms are $^{87}$Rb atoms, which are cooled and trapped by a MOT. After the atoms are released, they initially prepared in the $|g\rangle = |5S_{1/2}, F=2, m_g = 2\rangle$ state via optical pumping, where, $m$ denotes the magnetic quantum number. By applying write (pump) laser pulses on the $|g\rangle \to |e\rangle$ transition along x-axis (Figure 1b), we generate spontaneous Raman emissions of Stokes photons on the chiral ($\sigma^+$) transition $|e\rangle \leftrightarrow |s\rangle$ with $\Delta m = m_e - m_s = 1$, where, $|e\rangle = |5P_{1/2}, F'=2, m_e = 2\rangle$ is excited state, $|s\rangle = |5S_{1/2}, F=1, m_s = 1\rangle$ is a ground state, the write and Stokes transitions forms a three-level Λ configuration. The Stokes field propagating along the +z and –z -directions are coupled into the clock-wise (CW) and counter-clock-wise (CCW) modes of a running-wave cavity (shown in the following **Figure 2**), respectively. According to the selection rule [32], the $\sigma^+$ transition $\Delta m = 1$ effectively couples with the circularly-polarized light field $E = E_0 \frac{e_x + ie_y}{\sqrt{2}}$, where, $e_x$ ($e_y$) denotes the unite vector along x-axis (y-axis). For the Stokes fields emitting into +z (-z) -direction, it is rewritten as $E_k = E_0 \frac{e_x + ie_y}{\sqrt{2}} e^{-ikz}$ ( $E_{-k} = E_0 \frac{e_x + ie_y}{\sqrt{2}} e^{ikz}$ ), where, $k$ ($-k$) denotes their wave-vectors. In optics, the spins of the fields are related to their propagating direction. The forward-propagating field $E_k = E_0 \frac{e_x + ie_y}{\sqrt{2}} e^{-ikz}$ is right-hand circularly (RHC) polarization,[34] while the backward-propagating field $E_{-k} = E_0 \frac{e_x + ie_y}{\sqrt{2}} e^{ikz}$ is left-hand circularly (LHC) polarization, which can be intuitively seen from Figure 2.

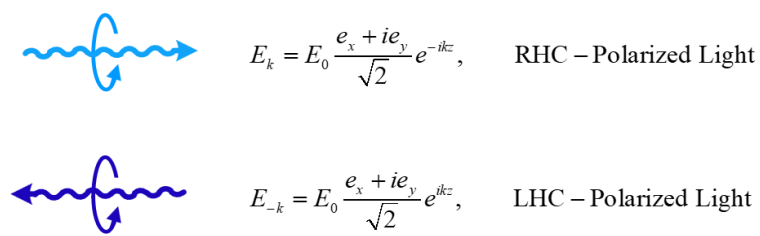

**Figure 2.** The relation between the expression and chirality as well as the propagation directions of circular polarization light.

We now explain that the spin-momentum correlation can be predicted by the reflection invariance, or left-right symmetric, of the atom-photon coupling system.[33] Based on the reflection invariance in atomic emission, the spin-momentum correlation has been theoretically predicted in Feynman lecture.[35] However, the quantum axis defined by a bias magnetic field is not involved in and then the magnetic quantum number $m$ of the atomic levels is ambiguous in the lecture. The upper in **Figure 3**(a) or Figure 3(b) shows the atoms in the excited state $|e\rangle = |5P_{1/2}, F'=2, m_e=2\rangle$. The bottom in Figure 3(a) is the original of Stokes emission by the transition $|e\rangle \rightarrow |s\rangle$ with $\Delta m_F = 1$, where, the Stokes field is along z-direction and is written as $E_k^{(R)} = E_0 \frac{e_x + ie_y}{\sqrt{2}} e^{-ikz}$, which is RHC-polarized. The bottom in Figure 3(b) is the mirror image of the Stokes emission by the transition $|e\rangle \rightarrow |s\rangle$, where, the Stokes field is along -z-direction and the direction of the external field $B_0$ is the same as that in the original due to that $B_0$ is axial vector.[35] Since the spins of the levels of the transition are defined by $B_0$, the transition $|e\rangle \rightarrow |s\rangle$ in the mirror imagine is the same as the origin and then has $\Delta m_F = 1$. The Stokes field components ($e_x$, $e_y$) are the polar vectors, whose signs are reversed,[35] they are changed into ($-e_x$, $-e_y$). So, the Stokes field in the Figure 3(b) may be written as $E_{-k}^{(L)} = E_0 \frac{e_x + ie_y}{\sqrt{2}} e^{ikz}$, which means that the backward-direction emission is left-hand circularly (LHC) polarized. Noted that the symbol "−" before $E_0$ in the expression of $E_{-k}^{(L)}$ is neglected since it is global and has not influence on the chirality of $E_{-k}^{(L)}$. The prediction for the spin-momentum correlation is not demonstrated in previous works.

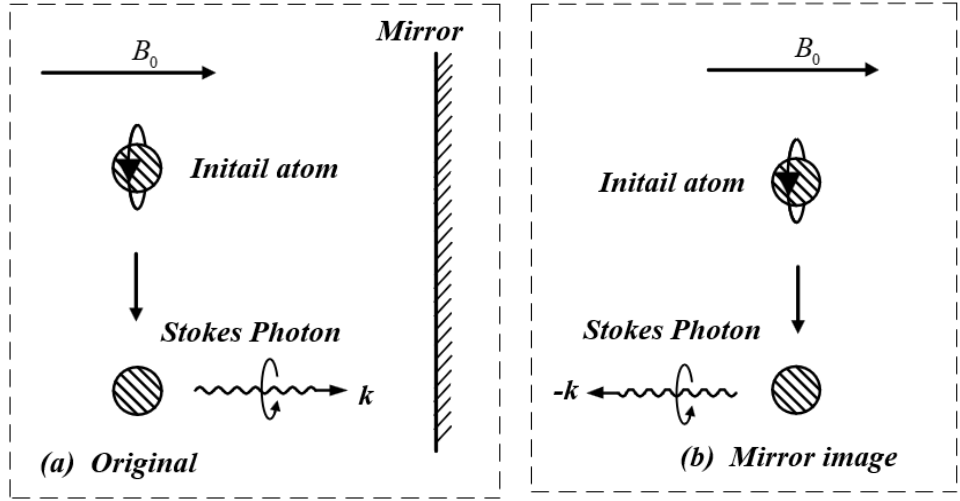

**Figure 3.** (a): The original of the Stokes emission from the atoms along z-direction. Figure 3. (b): The mirror image of the Stokes emission. The uppers in (a) and (b): the atoms are in the excited state. The bottom in (a) and (b): The Stokes emission by the atoms.

The application of the external magnetic field $B_0$ breaks the time-inverse symmetry of the atom-photon interaction system. For explain this, we firstly consider the case when the external bias magnetic field $B_0$ is not applied. In this case, the Hamilton $H_{a-p}$ of the atom coupling with the Stokes emission field is

$$H_{a-p} = \frac{1}{2m}\left(p - \frac{q}{c}A_p\right) + V(r) \qquad (1)$$

where, $p$ is the momentum operator of the electron in the individual atoms, $V(r)$ ($r$) is the Coulomb potential (position coordinate) of the electron in an individual atom, $A_p = A_{p0}e^{-i\omega_S t - ikz}$ is the vector potential of the Stokes field, $\omega_S$ is the frequency of the Stokes field. The interaction Hamilton $H_{a-p}$ for individual atoms and photons can be rewritten as

$$H_{a-p} = H_0 - \frac{q}{2mc} p \cdot A_0 + \frac{q^2}{2m^2c^2} A_p^2 \tag{2}$$

where, $H_0 = \frac{p^2}{2m} + V(r)$ is the Hamilton of the individual atoms, the term $\frac{q^2}{2m^2c^2} A_p^2$ is very small and neglected. We then have

$$H_{a-p} = H_0 - \frac{q}{2mc} p \cdot A_0 \tag{3}$$

The time reverse operation of Hamilton $H_{a-p}$ is

$$TH_{a-p}T^{-1} = TH_0 T^{-1} - \frac{q}{mc} T(pA_p) T^{-1} = H_0 - \frac{q}{mc}(-p)(-A_p) = H_{a-p}, \tag{4}$$

where, $TA_p T^{-1} = -A_p$ [36]. The above Eq. (4) shows that the atom-photon coupling at the absent the bias magnetic field is time-reverse symmetry. When the external bias magnetic field $B_0$ is applied, the Hamilton for the individual atom-photon systems is rewritten as

$$H'_{a-p} = \frac{1}{2m}\left(p - \frac{q}{c} A_P - \frac{q}{c} A_0\right)^2 + V(r)$$
$$\simeq H_0 - \frac{q}{2mc} pA_P - \frac{q}{2mc} pA_0 = H_{a-p} - \frac{q}{2mc} pA_0 \tag{5}$$

where, $A_0 = \frac{1}{2} B_0 \times r$ is the vector potential of the bias magnetic field $B_0$. The time reverse operation of Hamilton $H'_{a-p}$ is

$$TH'_{a-p} T^{-1} = TH_{a-p} T^{-1} - \frac{q}{2mc} TpT^{-1}TA_0 T^{-1} = H_{a-p} - \frac{q}{2mc}(-p)A_0 = H_{a-p} + \frac{q}{mc} PA_0 \neq H'_{a-p} \tag{6}$$

In the Eq.(6), $B_0$ ($A_0$) doesn't involve in the time inverse ($TA_0 T^{-1} = A_0$) since it is the external magnetic field. The result in Equation. (6) shows that atom-light system in the presence of $B_0$ is time-inverse asymmetry, meaning that the time-inverse symmetry is broke by the Bias field $B_0$. The time-inverse breaking induces that the coupling of an individual chiral atomic transition with circular-polarization light is non-reciprocal, which present a violation of the Kirchhoff's law of thermal radiation. We demonstrated this in the following.

## 3. Experimental results

Figure 3 is the sketch of the experimental set up, which includes a cold atomic ensemble that is coupled to a ring cavity inserted with a polarization interferometer (PI). With such a cavity, the RHC and LHC polarized light emissions will be distinguished. The ring cavity is formed by four flat mirrors M1, 2, 3 and OC, where, the M1, 2, 3 mirrors have high reflections, and the mirror OC is an output coupler that has a reflectance of 90%. The cold-atom ($^{87}$Rb) ensemble is inserted in a leg along z-axis of the cavity. The quantum axis of the atoms is defined by the bias of magnetic field $B_0$=12G along z-axis. Two quarter-wave plates $(\lambda/4)_1$ and $(\lambda/4)_2$ as well as two lenses (L1 and L2) are inserted at either end of the ensemble, respectively. The quarter-wave plates are used to change the circular polarizations of the emissions initially from the atoms into linear polarizations (more details see next Sec.) and the two lenses focus the cavity modes into a small spot size of 1mm at the center of the atoms. The two lenses focus the cavity modes into a small spot size of 1mm at the center of the atoms. The PI is formed by two beam displacers BD1 and BD2 [37]. The cavity finesse is F=17, and has a free spectrum range (FSR) and line bandwidth of 50MHz and ~3MHz, respectively. After the atoms are released from magneto-optical trap (MOT), they are optically prepared in the Zeeman state |s, $m_b = 2$>, and then we start the measurement trials.

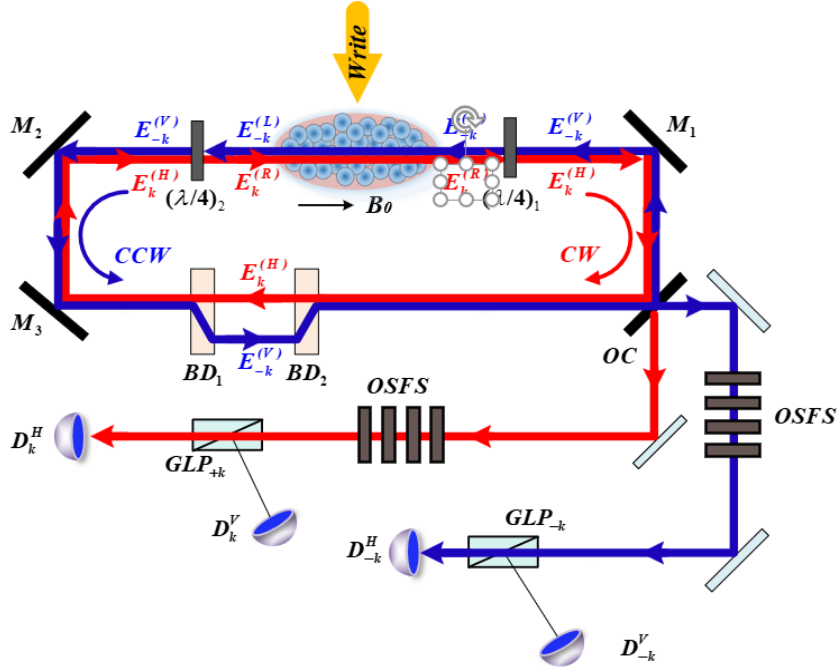

**Figure. 4** Experimental set up. The forward k and backward –k emissions are coupled into clockwise (blue line) and counter-clock-wise (black line) modes of the cavity, respectively. The labels RHC, LHC, H, and V denote right-hand-circular, left-hand-circular, horizontal and vertical polarizations of the Stokes emissions, respectively. GLP: Glan-Laser Prism; OSFS: Optical-Spectrum-Filter Set; BD1, BD2 : Beam Displacers; OC: Output coupler mirror; $D_k^H$, ($D_{-k}^H$) and $D_k^V$ ($D_{-k}^V$): Single photon detectors for detecting H-polarization and V-polarization components propagating along k (-k).

At the beginning of a trial, a 795-nm write laser pulse with a red-detuned by 110 MHz to the $|a, m_a = 2\rangle \rightarrow |e, m_e = 2\rangle$ transition is transversely applied onto the atoms (see Figure 1a). It is linear polarization paralleling with the z-axis. This write pulse induces spontaneous emissions of the Stokes photons on the chiral transition $|e, m_e = 2\rangle \rightarrow |b, m_b = 1\rangle$ ($\Delta m_F = 1$), which is DLCZ write process.[38] Emitting from the $\sigma^+$ transition, the Stoke field $E_k^{(R)}$ propagating along +z direction and the Stoke field $E_{-k}^{(L)}$ along -z

direction are coupled to the CW and CCW modes, respectively. As pointed out in the Sec. 2, the Stokes emissions from the atoms that propagate along z and –z directions are right-hand-circularly (RHC) and left-hand-circularly polarized, respectively, which results from the constraint to the spin-direction correction. They will go through the quarter-wave plates $(\lambda/4)_1$ and $(\lambda/4)_2$, respectively. The RHC polarized $E_k^{(R)}$ field is changed into H-polarized field labeled as $E_k^{(H)}$ after $(\lambda/4)_1$ and the LHC polarized $E_{-k}^{(L)}$ field into V-polarized field $E_{-k}^{(V)}$ after $(\lambda/4)_2$. The Jones matrixes of the wave plates $(\lambda/4)_1$ and $(\lambda/4)_2$ for +z and –z –direction polarization light are shown in Table 1. The fields $E_{+k}^{H}$ and $E_{-k}^{(V)}$ go through the polarization interferometer (PI), where, $E_{+k}^{H}$ passes through the arm 1 and $E_{-k}^{(V)}$ through arm 2 of the PI. Then, the $E_{+k}^{H}$ goes through $(\lambda/4)_2$ plate, whose polarization is changed back into the RHC polarization ($E_{+k}^{RHC}$), while $E_{-k}^{(V)}$ through $(\lambda/4)_1$ and whose polarization is changed back into LHC polarization ($E_{-k}^{LHC}$), both are coupled into the atoms again. The couplings of the atoms to the Stokes fields in the cavity are enhanced by a factor of $2F/\pi$ via Purcell effect. The polarization changes of Stokes fields via the quarter-wave plates according to Jones matrixes are explained in **Figure.S1** in Supplementary material.[39]

Table 1: The Jones matrixes M of the quarter wave plates for the cases that light propagate along forward (+z) and backward (-z) directions

| $(\lambda/4)_1$ | $M = \begin{pmatrix} 1 & -i \\ -i & 1 \end{pmatrix}$ | $M = \begin{pmatrix} 1 & i \\ i & 1 \end{pmatrix}$ |
|---|---|---|
| $(\lambda/4)_2$ | $M = \begin{pmatrix} 1 & i \\ i & 1 \end{pmatrix}$ | $M = \begin{pmatrix} 1 & -i \\ -i & 1 \end{pmatrix}$ |
|  | For the case that light propagate along forward (+z) | For the case that light propagate along backward (-z) |

For verifying the spin-momentum correlation, i.e., the polarizations of the forward (z-direction) and backward (-z-direction) scattering Stokes fields are RHC and LHC–polarized, respectively, we measure the polarization excitations of the Stokes photons at the two outputs of the cavity. The Stokes outputs along z and –z -axis from OC are directed into polarization beam splitters, i.e., Glan-Laser Prisms (GLPs), labeled as $GLP_{+k}$ and $GLP_{-k}$, respectively, which transmit $H$ polarization and reflect $V$ polarization light with high polarization extinction. The two outputs of $GLP_{+k}$ ($GLP_{-k}$) are sent to single-photon detectors $D_{+k}^{(H)}$ and $D_{+k}^{(V)}$ ($D_{-k}^{(H)}$ and $D_{-k}^{(V)}$), respectively.

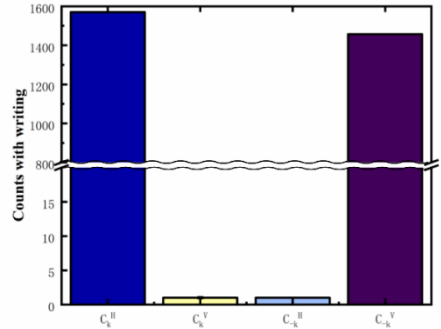

**Figure 5**. The measured histogram for $C_{k,-k}^{(H)}$, $C_{k,-k}^{(V)}$.

We use the ratio $\eta_{+k}^{(R/L)} = \gamma_{+k}^{(R)} / \gamma_{+k}^{(L)}$ ($\eta_{-k}^{(R/L)} = \gamma_{-k}^{(R)} / \gamma_{-k}^{(L)}$) to characterize the polarization extinction of the Stokes fields propagating along with z–direction (-z–direction), where, $\gamma_k^{(R)}$ ($\gamma_k^{(L)}$), $\gamma_{-k}^{(R)}$ ($\gamma_{-k}^{(L)}$) denote the spontaneous decay rates of the atoms emitting RHC (LHC) -polarization photons into the CW (CCW) mode. As mentioned above, the RHC-polarized (LHC-polarized) emission fields are transformed into the H (V) polarization. The H (V) -polarization outputs from the cavity are directed in to $D_k^{(H)}$ or $D_{-k}^{(H)}$, ($D_k^{(V)}$ or $D_{-k}^{(V)}$). So, in the presented experiment, the decay rates can be measured as: $\gamma_k^{(R)} \propto C_k^{(H)}$, $\gamma_k^{(L)} \propto C_k^{(V)}$, $\gamma_{-k}^{(R)} \propto C_{-k}^{(H)}$ and $\gamma_{-k}^{(L)} \propto C_{-k}^{(V)}$, where, $C_\alpha^{(X)}$ denotes the photon counting at the detectors $D_\alpha^{(X)}$ with $X = H,V$, and $\alpha = k, -k$. For an excitation probability $\chi_k = \chi_{-k} \sim 1\%$ of generating a Stokes photon along z and -z directions by a write pulse, we measured the cavity output by the detectors, the measured results of $C_{k,-k}^{(H)}$, $C_{k,-k}^{(V)}$, are plotted as histogram as shown in **Figure 5.** From it, we have

$$\gamma_k^{(R)} : \gamma_k^{(L)} : \gamma_{-k}^{(R)} : \gamma_{-k}^{(L)} \propto \left(C_k^{(H)} - N_{+k}^H\right) : \left(C_k^{(V)} - N_{+k}^{(V)}\right) : \left(C_{-k}^{(H)} - N_{-k}^H\right) : \left(C_{-k}^{(V)} - N_{-k}^V\right) \approx 1570:1:1:1458,$$

where, for example, $N_{+k}^{(H)}$ is the background noise for the channel of the detector $D_k^{(H)}$ without the write-pulse input. We then calculated the polarization extinction $\eta_{+k}^{(R/L)} = \frac{C_{+k}^{(H)} - N_{+k}^H}{C_{+k}^{(V)} - N_{+k}^{(V)}} \approx 1570:1$, $\eta_{-k}^{(R/L)} = \frac{\left(C_{+k}^{(H)} - N_{-k}^H\right)}{\left(C_{+k}^{(V)} - N_{-k}^V\right)} \approx 1:1458$, respectively. The measured results show that the Stokes fields propagating along z and –z directions are RHC and LHC polarizations, respectively, which is in agreement with the prediction of the spin-momentum correlation mentioned above. This spin-momentum correlation means that the coupling between the Stokes emissions the chiral atoms is nonreciprocal, which origin from temporal symmetries (Lorentz reciprocity) breaking by the bias magnetic field.[28-31] Such nonreciprocal couplings of a circular-polarization field with the atoms show a violation of Kirchhoff's law of thermal radiation. Specifically, the Kirchhoff's law describe that the absorptivity ($\alpha$) and emissivity ($e$) of a body at a given temperature ($T$) and a wavelength ($\lambda$), angle ($\theta$) and polarization ($\mu$) are equal, i.e.,

$$\alpha(T,\lambda,\theta)_\mu = e(T,\lambda,\theta)_\mu \qquad (7).$$

With the pumping by a weak writing laser (~0.4 mW/ cm$^2$) in the presented experiment, the temperature of the cold atomic ensemble will be raised by a little value and then is slightly larger than that (100 μk) of the cold atoms. For the coupling of the RHC (LHC) polarized Stokes fields, the emissive $e(T,\lambda,\theta=0^0)_R \propto \gamma_{+k}^{(R)}$ ( $e(T,\lambda,\theta=0^0)_L \propto \gamma_{+k}^{(L)}$ ), while the absorptivity $\alpha(T,\lambda,\theta=0^0)_R \propto \gamma_{-k}^{(R)}$ ($\alpha(T,\lambda,\theta=0^0)_L \propto \gamma_{-k}^{(L)}$). Our measured result shows that $\gamma_k^{(R)} \gg \gamma_{-k}^{(R)}$ ( $\gamma_k^{(L)} \ll \gamma_{-k}^{(L)}$ ), means that $e(T,\lambda,\theta=0^0)_R \gg \alpha(T,\lambda,\theta=0^0)_R$ $e(T,\lambda,\theta=0^0)_L \ll \alpha(T,\lambda,\theta=0^0)_L$.

Based on the spin-momentum correlation, we demonstrated unidirectional Raman emissions of Stokes photons in the ring cavity via dissipating V- or H- polarization field in the ring cavity. By setting an optical blocker in the arm 1 and 2 of the PI as shown in Figure 6 (a) and (b), respectively, we dissipate the V- and H- polarization mode and demonstrated the forward- or backward- direction Stokes emission, respectively. When the H-polarized CW mode is blocked (Figure 6 (a)), the measured outputs of $GLP_{+k}$, $GLP_{-k}$ by the detectors $D_{+k}^{(H)}$, $D_{+k}^{(V)}$, $D_{-k}^{(H)}$ and $D_{-k}^{(V)}$ are

$$\gamma_k^{(R)}:\gamma_k^{(L)}:\gamma_{-k}^{(L)}:\gamma_{-k}^{(R)} \approx \left(C_{+k}^{(H)}-N\right):\left(C_{+k}^{(V)}-N\right):\left(C_{-k}^{(V)}-N\right):\left(C_{-k}^{(H)}-N\right) \approx 1570:1:1:1 \quad ,$$

while, when the V-polarized CCW mode is blocked (Figure 5b), the measured outputs are

$$\gamma_k^{(R)}:\gamma_k^{(L)}:\gamma_{-k}^{(L)}:\gamma_{-k}^{(R)} \approx \left(C_{+k}^{(H)}-N\right):\left(C_{+k}^{(V)}-N\right):\left(C_{-k}^{(V)}-N\right):\left(C_{-k}^{(H)}-N\right) \approx 1:1:1:1458 .$$

The interaction between the atoms and the pair of counterpropagating modes can be described by the directional $\beta$ factor, which is defined as the ratio of the spontaneous emission rate into the $\pm k$ mode pair and the total emission rate [1]: $\beta_\pm = \gamma_{\pm k}/(\gamma_k + \gamma_{-k} + \Gamma)$, where, $\Gamma$ is the emission rate into all other modes. Since the cavity in the presented experiment has a low finesse (~17), the emission rate $\Gamma$ is far beyond that into the $\pm k$ modes ($\Gamma \gg \gamma_{\pm k}$), the directional $\beta$ factor is closed to zero in the presented experiment. When using a high-finesse cavity instead of the presented cavity, the strong atom-photon coupling will make the spontaneous emission rate $\Gamma$ to be closed to zero. In this case, one can obtain $\beta_+ \to 1$ and $\beta_- \to 0$ (or $\beta_- \to 1$ and $\beta_+ \to 0$), i.e., ideal unidirectional emission. For the presented experiment, we define the relative directional factor, which is written as $\beta_\pm' = \beta_\pm / \beta_\mp = \gamma_{\pm k}/\gamma_{\mp k}$, to describe the ratio of the spontaneous emission rate into the $\pm k$ mode pair. In Figure 6(a) we obtained $\beta_+' \approx 99.87\%$, while, in Figure 6(b), we $\beta_-' \approx 99.87\%$, which clearly show the perfect unidirectional emission in the chiral atom-cavity system.

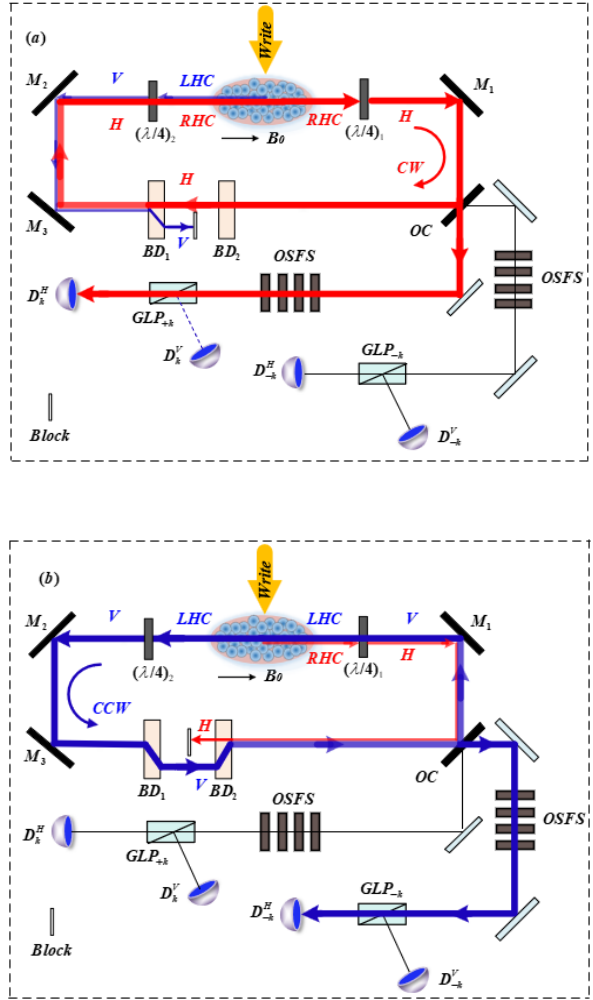

**Figure. 6** (a): The experimental setup for the case that the CCW mode is blocked (a), and the experimental setup for that the CW mode is blocked (b). The labels are the same as that in Figure 4.

## 4. Conclusion

Based on the left-right-symmetry, i.e., reflection invariance [33] in the atom-photon couplings, we pointed out that the spins (circular polarization) of the Stokes fields from a chiral atomic transition are correlated with their propagation along z and –z directions. Certainly, for achieving such spin-direction correction, we have to apply a bias magnetic field $B_0$ on the atoms along z-direction to define the quantum axis, which break the time inverse symmetry of the atom-light interface. We noted that by applying strong magnetic fields on optical material, people break Lorentz time symmetry of light propagation in optical isolators based on Faraday magneto-optical effect,[1] where, an individual isolator promises an input light beam to propagate in z- direction and prevent the beam from going back the forward path. Recently, the violation of Kirchhoff's law of thermal radiation has been demonstrated with magneto-optic photonic structure materials.[28-30] So far, the Kirchhoff's violation has not been demonstrated in atomic vapors with a weak magnetic field applied on them. In the presented work, we experimentally demonstrated the spin-momentum correlation of the Stokes emissions from the chiral transition of the atoms. Constrained to spin-momentum correlation, we observed that the Stokes emissions present a violation of the Kirchhoff's law of thermal radiation. The measured polarization extinctions of the emissions along forward- or backward-direction are up to ~1500:1, which represent a very high precision demonstration of the conversion of angular momentum conservation of atomic emission process to our knowledge. The presented demonstration of the unidirectional Raman emissions of Stokes photons in the ring cavity can be found application in chiral quantum networks,[2] in directional phase-sensitive amplifier,[40] or suppressing backward spontaneous Raman scattering of Stokes photons in a standing-wave cavity.[41] Based on the spin-momentum correlation, we design an experimental setup including linear optical elements (see supplementary material [39]), which promise the interaction between two atoms in free space to be non-reciprocity. Such non-reciprocity interaction of atoms can be used for some quantum optical tasks[42-43] and simulations of many body physics.[5]

## Acknowledgements

We thank the National Natural Science Foundation of China (12174235), the Fund for Shanxi Key Subjects Construction (1331), and the Fundamental Research Program of Shanxi Province (202203021221011).

*Corresponding author: wanghai@sxu.edu.cn

Keywords: non-reciprocity interaction between atoms and photons; spin-momentum correlation; unidirectional Raman emission;

violation of Kirchhoff law

# Supplementary material

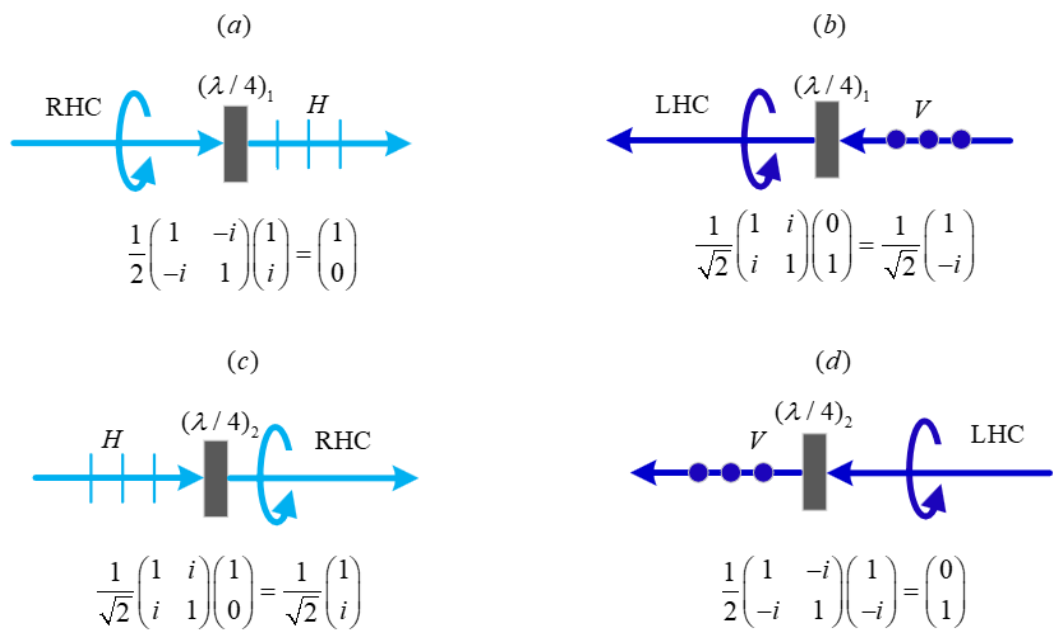

**Figure S1.** The polarization changes of Stokes fields via the quarter-wave plates according to Jones matrixes.

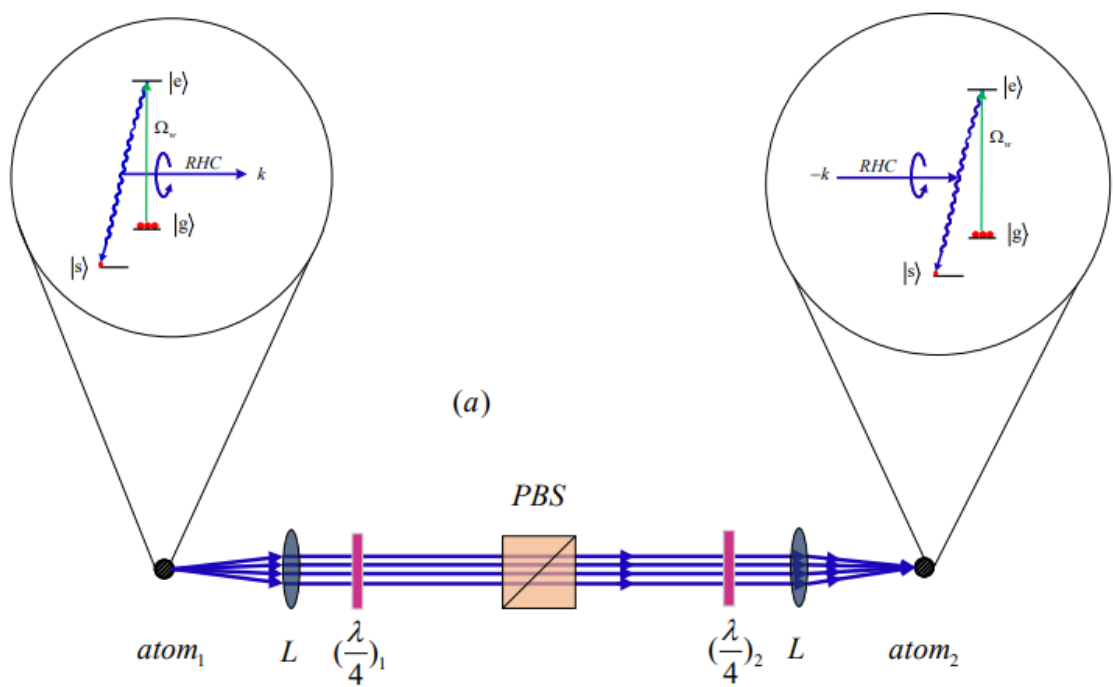

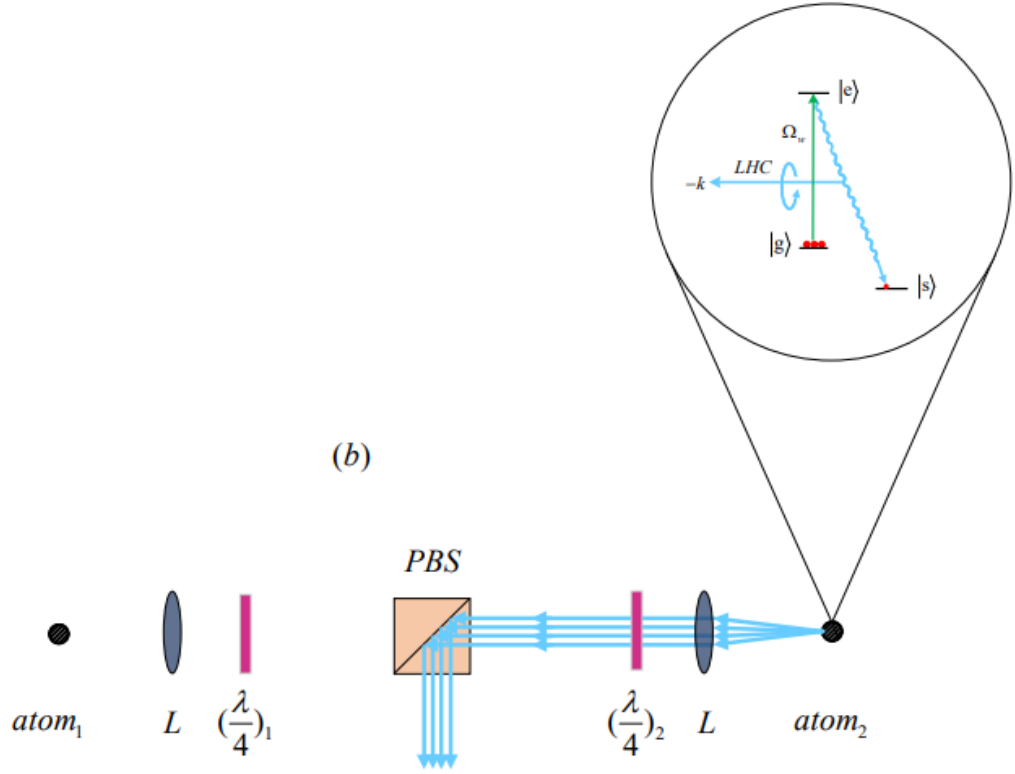

**Figure S2.** The non-reciprocity interaction between two atoms. (a) Stokes photons emitted from atom1, it is RHC polarized. Its polarization is changed into H-polarization after $(\frac{\lambda}{4})_1$ and then pass through the polarization beam splitter (PBS). It then changed into RHC- polarization and is interacted with the atom2. (b) Stokes photons emitted from atom2, it is LHC polarized. Its polarization is changed into V-polarization by a $(\frac{\lambda}{4})_2$ and then reflected by the polarization beam splitter (PBS). It then can't seen by the atom1. $\Omega_w$: write (pump) laser pulses on the transition. L: lens.